\documentclass[aps,amsfonts,amsmath,prd,preprint,nofootinbib]{revtex4}
\usepackage{epsf}

\newcommand{\beq}{\begin{equation}}
\newcommand{\eeq}{\end{equation}}

\begin{document}

\title{A measure of the multiverse}

\author{Alexander Vilenkin}

\address{ Institute of Cosmology, Department of Physics and Astronomy,\\ 
Tufts University, Medford, MA 02155, USA}

\begin{abstract}

I review recent progress in defining a probability measure in the
inflationary multiverse. General requirements for a satisfactory
measure are formulated and recent proposals for the measure are
clarified and discussed.
  
\end{abstract}

\maketitle

\section{Introduction}

String theory appears to have a multitude of solutions describing
vacua with different values of the low-energy constants. The number of
vacua in this vast ``landscape'' of possibilities can be as large as
$10^{500}$ \cite{BP,Susskind,Douglas}. In the cosmological context,
high-energy vacua drive exponential inflationary expansion of the
universe. Transitions between different vacua occur through tunneling
and quantum diffusion, with regions of different vacua nucleating and
expanding in the never-ending process of eternal inflation
\cite{AV83,Linde86}. As a result, the entire landscape
of vacua is explored.

If indeed this kind of picture describes our universe, we will
never be able to calculate all constants of Nature from first
principles. At best we may only be able to make statistical
predictions. The key problem is then to calculate the probability
distribution for the constants. It is often referred to as {\it the
measure problem}.

The probability $P_j$ of observing vacuum $j$ can be expressed as a
product 
\beq
P_j=P_j^{(prior)} f_j
\label{Pj}
\eeq
where the prior probability $P_j^{(prior)}$ is determined by the
geography of the landscape and by the dynamics of eternal inflation,
and the selection factor $f_j$ characterizes the chances for an
observer to evolve in vacuum $j$. The distribution (\ref{Pj}) gives
the probability for a randomly picked observer to be in a given
vacuum.

It seems natural to identify the prior probability with the fraction
of volume $P_j^{(V)}$ occupied by a given vacuum and the selection
factor with the number of observers $n_j^{(obs)}$ per unit
volume\footnote{The product in (\ref{Pj}) should of course be properly
normalized.} \cite{AV95}, 
\beq
P_j^{(prior)}\propto P_j^{(V)},
\label{old1}
\eeq
\beq
f_j \propto n_j^{(obs)}.
\label{old2}
\eeq
This approach, however, encounters a severe difficulty: the result
sensitively depends on the choice of a spacelike hypersurface (a
constant-time surface) on which the distribution is to be evaluated.
This problem was uncovered by Andrei Linde and his collaborators when
they first attempted to calculate volume distributions
\cite{LLM,LM,GBL}. It eluded resolution for more than a decade, but
recently there have been some promising developments, and I believe we
are getting close to completely solving the problem. Here, I will
briefly discuss the nature of the dificulty and then review the new
proposals for $P_j$. Most of this discussion is based on my work with
Jaume Garriga, Delia Schwartz-Perlov, Vitaly Vanchurin, and Serge
Winitzki \cite{GSPVW,Vitaly} (see also \cite{landscape1}).

\section{Problem with global-time measure}

The spacetime structure of an eternally inflating universe is
schematically illustrated in Fig.1. For simplicity, we shall focus on
the case where transitions between different vacua occur only through
bubble nucleation.  The bubbles expand rapidly approaching the speed
of light, so their worldsheets are well approximated by light
cones. Disregarding quantum fluctuations, bubble interiors are open
FRW universes \cite{CdL}; they are often called ``pocket
universes''.  If the vacuum inside a bubble has positive energy
density, it becomes a site of further bubble nucleation; we call such
vacua ``recyclable''.  Negative-energy vacua, on the other hand,
quickly develop curvature singularities; we shall call them ``terminal
vacua''.

\begin{figure}
\begin{center}
\leavevmode\epsfxsize=5in\epsfbox{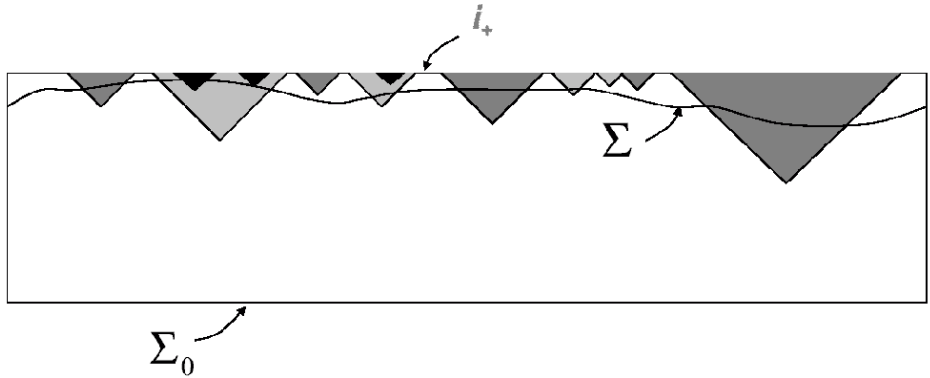}
\caption{A schematic conformal diagram for a comoving region in an 
eternally inflating universe. Bubbles of different vacua are represented 
by different shades of gray. The upper boundary of the diagram $i_+$ is the 
future timelike infinity. A surface of constant global time $\Sigma$ cuts 
through the entire region and intersects many bubbles.} 
\end{center}
\end{figure}

The diagram represents a comoving region, which is initially
comparable to the horizon. The initial moment is a spacelike
hypersurface $\Sigma_0$, represented by the lower horizontal boundary
of the diagram, while the upper boundary represents future infinity,
when the region and all the bubbles become infinitely large.
How can we find the fraction of volume occupied by different vacua? A
natural thing to do is to consider a spacelike hypersurface $\Sigma$,
which cuts through the entire region, as shown in the figure. If $t$
is a globally defined time coordinate, then all surfaces $t={\rm
const}$ will have this property. One can use, for example, the proper
time along the ``comoving'' geodesics orthogonal to the surface
$\Sigma_0$.\footnote{The term ``comoving'' is used very loosely here,
since the vacuum does not define any rest frame. Any congruence of
geodesics orthogonal to a smooth spacelike surface
$\Sigma_0$ can be regarded as ``comoving''.}  Alternatively, one could
use the so-called scale factor time, defined as a logarithm of the
expansion factor along the comoving geodesics, or any other suitable
time coordinate. Once the time coordinate is specified, one can find
the fraction of volume occupied by different vacua on the surface
$t={\rm const}$ and then take the limit $t\to\infty$.

Unfortunately, as I have already mentioned, the result of this
calculation is sensitively dependent on one's choice of the time
coordinate \cite{LLM}. The reason is that the volume of an eternally
inflating universe is growing exponentially with time. The volumes of
regions filled with all possible vacua are growing exponentially as
well. At any time, a substantial part of the total volume is in new
bubbles which have just nucleated. Which of these bubbles are cut by
the surface depends on how the surface is drawn; hence the
gauge-dependence of the result. Since time is an arbitrary label in
General Relativity, none of the possible choices of the global time
coordinate appears to be preferred.  For more discussion of this
gauge-dependence problem, see \cite{Guth,Tegmark,Winitzki1}.

\section{General requirements}

At this point, it will be useful to formulate some general
requirements that any satisfactory definition of $P_j$ should comply
with \cite{GSPVW}. 

First of all, we require that $P_j$ should not depend on gauge, that
is, on arbitrary choice of a hypersurface.  More generally, it should
not depend on any arbitrary choices.

Second, we require that $P_j$ should be independent of the initial
conditions at the onset of inflation. The dynamics of eternal
inflation is an attractor; its asymptotic behavior has no memory of
the initial state.\footnote{I assume that any vacuum is accessible
through bubble nucleation from any other vacuum. Alternatively, if the
landscape splits into several disconnected domains which cannot be
accessed from one another, each domain will be characterized by an
independent probability distribution.} We believe that the
probabilities should also have this property. Note that this
condition is not satisfied by an earlier proposal in \cite{markers} and by
a more recent proposal in \cite{Bousso06}.

\section{A pocket-based measure}

\subsection{Bubble abundance $p_j$}

We shall now discuss the new proposal for $P_j$, introduced in Ref. 
\cite{GSPVW}. The presentation here is somewhat different from 
\cite{GSPVW}, but the essence is the same. 

The idea is that instead of trying to compare volumes occupied by
different vacua, we compare the numbers of different types of bubbles (pocket
universes). Thus, instead of Eq.~(\ref{old1}), we make the assignment
\beq
P_j^{(prior)}\propto p_j,
\label{Pp}
\eeq
where $p_j$ is the abundance of $j$-type bubbles. (We shall see later
that the volume expansion in this approach is accounted for in the
selection factor $f_j$.)

The definition of $p_j$ is a tricky business, because the total number
of bubbles is infinite, even in a region of a finite comoving
size.\footnote{The problem of calculating $p_j$ is somewhat similar to
the question of what fraction of all natural numbers are odd. The
answer depends on how the numbers are ordered. With the standard
ordering, $1,2,3,4, ...$, the fraction of odd numbers in a long
stretch of the sequence is 1/2, but if one uses an alternative
ordering $1,3,~2,~5,7,~4, ...$, the result would be 2/3. One could
argue that, in the case of integers, the standard ordering is more
natural, so the correct answer is 1/2. Here we seek an analogous
ordering criterion for the bubbles.} We thus need to introduce some
sort of a cutoff. 

The proposal of \cite{GSPVW} is very simple: count only bubbles
greater than a certain comoving size $\epsilon$, and then take the
limit $\epsilon\to 0$. That is,
\beq
p_j = {\rm lim}_{\epsilon\to 0} {N_j(>\epsilon)\over{N(>\epsilon)}}.
\label{pjNj}
\eeq
To define the comoving size, one has to specify a congruence of
``comoving'' geodesics emanating (orthogonally) from some initial
spacelike hypersurface $\Sigma_0$. As they extend to the future, the
geodesics will generally cross a number of bubbles before ending up in
one of the terminal bubbles, where inflation comes to an end. There
will also be a (measure zero) set of geodesics which never hit
terminal bubbles. The starting points of these geodesics on $\Sigma_0$
provide a mapping of the eternally inflating fractal
\cite{Aryal,Winitzki,Winitzki2}, consisting of points on $i_+$ where
inflation never ends. In the same manner, each bubble encountered by
the geodesics will also be mapped on $\Sigma_0$, and we can define the
comoving size of a bubble as the volume of its image on
$\Sigma_0$. (The volume of a bubble is calculated including all the
daughter bubbles that nucleate within it.) Throughout this paper, we
disregard bubble collisions.

I will now argue that the above prescription satisfies the
requirements formulated in Section III. 

In an inflating spacetime, geodesics are rapidly diverging, so bubbles
formed at later times have a smaller comoving size. (The comoving size
of a bubble is set by the horizon at the time of bubble nucleation.)
The bubble counting can be done in an arbitrarily small neighborhood
$\delta$ of any point belonging to the ``eternal fractal'' image on
$\Sigma_0$.  Every such neighborhood (except a set of relative measure
zero) will contain an infinite number of bubbles of all kinds and will
be dominated by bubbles formed at very late times and having very
small comoving sizes. The resulting values of $p_j$, obtained in the
limit of bubble size $\epsilon\to 0$, will be the same in all such
neighborhoods, because of the universal asymptotic behavior of eternal
inflation. The same result will also hold in any finite-size region on
$\Sigma_0$ (provided that it contains at least one ``eternal point'').

The values of $p_j$ are independent of the choice of the initial
hypersurface $\Sigma_0$. Once again, this is a consequence of the
universal, attractor behavior of eternal inflation. Mathematically,
this is reflected in the fact that the asymptotic distributions
obtained from the Fokker-Planck equation (in the case of slow-roll
models \cite{AV83,Starobinsky,LLM}) and from the master equation (in
the case of bubble nucleation models \cite{recycling}) do not depend
on the initial surface that was used to define the comoving
congruence.\footnote{Another way to see this is to consider a small
patch of $\Sigma_0$ including an eternal point. If the patch is small
enough, it can be regarded as flat. Then any change of $\Sigma_0$ will
amount to changing the orientation of its normal, that is, the
4-velocity of the ``comoving'' congruence. But since there is no
preferred frame in the inflating de Sitter-like spacetime, all choices
will result in the same asymptotic behavior and will yield identical
values of $p_j$.}

The condition of orthogonality between the congruence and the
hypersurface $\Sigma_0$ can be relaxed. Suppose we change $\Sigma_0$
while keeping the congruence fixed, so that the congruence and
$\Sigma_0$ are no longer orthogonal. Once again, focusing on the
vicinity of an eternal point, any change of the hypersurface amounts to a
constant rescaling of all bubble sizes and has no effect on $p_j$.

Moreover, although we use the metric on $\Sigma_0$ to compare the
bubble sizes, the results are unaffected by arbitrary smooth 
transformations of the metric.  Any such transformation will locally
be seen as a linear transformation, which amounts to a constant
rescaling. In a sufficiently small patch of $\Sigma_0$, all bubble
volumes are rescaled in the same way, so the bubble counting should not
be affected.

The results obtained using this method are also independent of the
initial conditions at the onset of eternal inflation.  This simply
follows from the facts that the bubble counting is dominated by late
times and that the asymptotic behavior in eternal inflation is
independent of the initial state.

The calculation of bubble abundances, defined by Eq.(\ref{pjNj}), can
be reduced to an eigenvalue problem for a matrix constructed out of
the transition rates between different vacua
\cite{GSPVW}.\footnote{The calculation in \cite{GSPVW} assumes that
the divergence of geodesics is everywhere determined by the local
vacuum energy density. This is somewhat inaccurate, since it ignores
the brief transition periods following the bubble crossings and the
focusing effect of the domain walls. The accuracy of the method is
expected to be up to factors $O(1)$. A more detailed discussion will
be given elsewhere \cite{walls}.} This prescription has been tried on
some simple models and appears to give reasonable results
\cite{GSPVW,SPV}. For example, if there is a single false vacuum,
which can decay into a number of vacua with nucleation rates
$\Gamma_j$ , one finds
\beq
p_j\propto \Gamma_j,
\eeq
as intuitively expected.

\subsection{Eternal observers}

Another interesting special case is that of full recycling. If all
vacua have positive energy density, $\rho_j >0$, there are no
terminal vacua and all geodesics of the congruence represent ``eternal
observers'', who endlessly transit from one vacuum to another,
exploring the entire landscape. The distribution $p_j$ in this case
can be found in a closed form \cite{Vitaly}:
\beq
p_j \propto \sum_i \Gamma_{ji}e^{S_i},
\label{eternal}
\eeq
where 
$\Gamma_{ji}$ is the nucleation rate of $j$-type bubbles in
vacuum $i$, $S_i=\pi/H_i^2$ is the Gibbons-Hawking entropy of de
Sitter space, and 
\beq
H_j=(8\pi G\rho_j/3)^{1/2}
\label{Hj}
\eeq
is the expansion rate corresponding to the local vacuum energy density
$\rho_j$.
 
In the case of full recycling, the bubble abundance $p_j$ can also be
defined as the frequency at which $j$-type bubbles are visited along
the worldline of an eternal observer. This definition involves
observations accessible to a single observer - a property that some
string theorists find desirable \cite{Susskind,Bousso06}. It has been shown in
\cite{Vitaly} that this eternal-observer definition gives the same 
result (\ref{eternal}) as the pocket-based measure of \cite{GSPVW}. 

For example, in the simplest case of only two vacua,
Eq.~(\ref{eternal}) gives $p_1\propto \Gamma_{12}e^{S_2}$, $p_2\propto
\Gamma_{21} e^{S_1}$, and using the property \cite{EWeinberg} 
\beq
\Gamma_{ij}/\Gamma_{ji}=e^{S_i - S_j},
\eeq
we obtain $p_1=p_2=0.5$. This is, of course, in agreement with the
frequency of visiting the two vacua: the frequency should be the same,
since the eternal observer goes back and forth between them.

\subsection{An equivalent proposal}

An alternative prescription for $p_j$ has been suggested by Easther,
Lim and Martin \cite{ELM}. They randomly select a large number ${\cal
N}$ of points on a compact patch of a spacelike hypersurface
$\Sigma_0$ in the inflating part of spacetime. They follow the
geodesics emanating from these points and check which bubbles they
cross. The bubble abundance is then defined as
\beq
p_j = {\rm lim}_{{\cal N}\to\infty}{{\cal N}_j\over{\cal N}},
\label{Easther}
\eeq
where ${\cal N}_j$ is the number of type-$j$ bubbles crossed by at
least one geodesic.

As the number of points is increased, the average distance $\delta$
between them on $\Sigma_0$ gets smaller, so most bubbles of comoving
volume larger than $\epsilon\sim \delta^3$ are counted.  In the limit
of ${\cal N}\to\infty$, we have $\epsilon\to 0$, and it is not difficult to
see that this prescription is equivalent to the one described in the
preceding subsection. (For a rigorous proof, see {\it Note added} in
\cite{GSPVW}.)

Easther et. al. argue that the values of $p_j$ in (\ref{Easther}) are
independent of the choice of measure on $\Sigma_0$. This is consistent
with our analysis. They also argue that the initial velocities of the
worldlines on $\Sigma_0$ can be chosen at random without affecting the
$p_j$. I think this statement needs to be modified. If the velocities
at neighboring points are chosen independently, then in the limit
$n \to\infty$ the velocity distribution on $\Sigma_0$ will be
very singular, and I see no reason to expect that $p_j$ will remain
unchanged. On the other hand, we do expect $p_j$ to be invariant under
continuous variations of the geodesic congruence, as explained in
Section IV A.

\subsection{Bousso's proposal}

Raphael Bousso \cite{Bousso06} (see also \cite{BFY06}) suggested an
extension of the prescription in \cite{Vitaly} to the case when there
are some terminal bubbles, so the observers are generally not
eternal. The idea is to start with an ensemble of observers
characterized by some initial distribution. All observers, except a
set of measure zero, will end up in terminal bubbles after visiting a
certain number of recyclable bubbles. Bousso's proposal is that the
measure $p_j$ should be proportional to the total number of times the
observers in the ensemble visit bubbles of type $j$.

The resulting measure is strongly dependent on the initial
distribution function, so one has to address the question of where
that distribution comes from. Bousso suggests it might be derived from
the wave function of the universe $\Psi$. The usual interpretation of
$\Psi$ is that it gives probabilities for different initial states as
the universe nucleates out of nothing. The nucleation is followed by
eternal inflation, which produces an unlimited number of all possible
bubbles, so the initial state is quickly forgotten. Bousso's proposal
is based on a very different, holographic view, which asserts that the
region outside the horizon should be completely excluded from
consideration. Hence, one is dealing with an ensemble of disconnected
horizon-size regions nucleating out of nothing. For someone not
initiated in holography, this view is very hard to adopt, but as long
as it is mathematically consistent, one can work out its predictions
and compare them with the data.

\section{The selection factor $f_j$}

The selection factor $f_j$ should characterize the relative number of
observers in different types of pockets. As I already mentioned, the
interior spacetime of a pocket is that of an open FRW universe, so
each pocket that has any observers in it has an infinite number of
them. In order to compare the numbers of observers, we will have to
define a comoving length scale $R_j$ on which observers are to be
counted in bubbles of type $j$. The first thing that comes to mind is
to set $R_j$ to be the same for all bubbles.  However, this is not
enough. The expansion rate is different in different bubbles, so the
physical length scales corresponding to $R_j$ will not stay equal,
even if they were equal at some moment. We could specify the times
$t_j$ at which $R_j$ are set to be equal, but any such choice would be
subject to the criticism of being arbitrary.

A possible way around this difficulty was proposed in \cite{GSPVW}. At
early times after nucleation, the dynamics of open FRW universes
inside bubbles is dominated by the curvature, with the scale factor
given by
\beq
a_j(t)\approx t
\label{at}
\eeq
for all types of bubbles. For example, for a quasi-de Sitter bubble
interior,
\beq
a_j(t)\approx H_j^{-1}\sinh (H_j t),
\eeq
where $H_j$ is given by Eq.~(\ref{Hj}).  The specific form of the
scale factor at late times is not important for our argument. The
point is that for $t\ll H_j^{-1}$ all bubble spacetimes are nearly
identical, with the scale factor (\ref{at}).

The proposal of \cite{GSPVW} is that the reference scales should be
chosen so that $R_j$ are the same at some small $t=\tau$ (same for all
bubbles). The choice of $\tau$ is unimportant, as long as $\tau\ll
H_j^{-1}$ for all $j$. Then, up to a constant, the physical length
corresponding to $R_j$ is
\beq
R_j^{(phys)}(t)=a_j(t).
\label{Ra}
\eeq
For times $t\gg H_j^{-1}$, this can be expressed as 
\beq
R_j^{(phys)}(t)\approx H_j^{-1}Z_j(t),
\label{RZ}
\eeq
where $Z_j$ is the expansion factor since the onset of the
inflationary expansion inside the bubble ($t\sim H_j^{-1}$). 

Alternatively, $R_j^{(phys)}$ in (\ref{RZ}) can be identified as the
curvature scale. It is the characteristic large-scale curvature radius
of the bubble universe. This definition makes no reference to early
times close to the bubble nucleation: the curvature radius can be
found at any time. It is, in principle, a measurable quantity.

The selection factor $f_j$ can thus be written as
\beq
f_j \propto n_j,
\label{fn}
\eeq
where $n_j$ is the number of observers who will evolve per unit
comoving volume (normalized at the same $\tau \ll H_j^{-1}$ for all
bubbles). The calculation of $n_j$ is of course a challenging
problem; I will not address it here.

It follows from Eqs.~(\ref{RZ}) and (\ref{fn}) that large inflation
inside bubbles is rewarded with our definition of the measure. An
inflationary expansion by a factor $Z$ enhances the probability by
$Z^3$.

\section{Continuous variables}

Our prescription for the measure can be straightforwardly generalized
to the case when, in addition to bubbles, there are some continuously
varying fields $X$. Eq.~(\ref{Pp}) for the prior is replaced by
\beq
P_j^{(prior)}\propto  p_j {\hat P}_j(X),
\label{hatP}
\eeq
where ${\hat P}_j(X)$ is the normalized distribution for $X$ in a
bubble of type $j$ at $t=\tau \ll H_j^{-1}$,
\beq
\int{\hat P}_j(X)dX=1.
\eeq
This distribution is determined by the dynamics of quantum fields $X$
during inflation. It can be calculated analytically or numerically,
using the methods of Refs.~\cite{VVW,GSPVW}.

Eq.~(\ref{fn}) for the selection factor is replaced by
\beq
f_j(X)\propto n_j(X).
\eeq

\section{Discussion}

The above definition of the measure is just a proposal. We have not
derived it from first principles. In fact, there is no guarantee that
there is some unique measure that can be used for making predictions
in the multiverse. How, then, can we ever know that we made the right
choice out of all possible options?

What I find encouraging is that even a single definition of measure
that satisfies some basic requirements proved very difficult to find.
It is also reassuring that alternative prescriptions suggested in
\cite{ELM} and \cite{Vitaly} turned out to be equivalent to the 
pocket-based measure of \cite{GSPVW}. 

Here, we required that the measure should not depend on any arbitrary
choices, such as the choice of gauge or of a spacelike hypersurface,
and that it should be independent of the initial conditions at the
onset of inflation. These conditions, however, do not specify the
measure uniquely. For example, a flat measure, $p_j={\rm const}$ for
all $j$, clearly satisfies the conditions. It would be interesting to
formulate a set of requirements which selects a unique definition of
the measure. 

Bubbles of different types can generally collide, with domain walls
forming to separate the different vacua. Our prescription for the
measure needs to be generalized to include these processes. Another
necessary extension is to the case where transitions betwen vacua can
occur through quantum diffusion. (Some steps in this direction have
been made in \cite{GSPVW}.)

The ultimate test of any proposed measure will be a comparison of its
predictions with observations. The first attempts in this direction
have already produced some intriguing results
\cite{Aguirre,Freivogel,Hall1,QLambda,Tegmark2,Hall2,SPV,Hawking}.

~~~~~~~~~~~~~~~~~~~

I am grateful to Raphael Bousso, Jaume Garriga, Alan Guth, Delia
Schwartz-Perlov, Leonard Susskind, Vitaly Vanchurin and Serge Winitzki
for discussions and useful comments.  This work was supported in part
by the Foundational Questions Institute and by the National Science
Foundation.

\end{document}